\documentclass[pre,twocolumn,10pt,floatfix]{revtex4-1}
 \usepackage[utf8x]{inputenc}
 \usepackage[T1]{fontenc}
\usepackage{diagbox}
\usepackage{amsfonts}
\usepackage{amsmath}
\usepackage{amsthm}
\usepackage{indentfirst}
\usepackage{microtype}
\usepackage[squaren]{SIunits}
\usepackage{booktabs}
\usepackage{graphicx}
\usepackage{multirow}
\usepackage{wallpaper}

\usepackage{braket}
\usepackage{caption}
\usepackage{subcaption}
\newcommand{\beq}{\begin{eqnarray}}
\newcommand{\eeq}{\end{eqnarray}}
\newcommand{\be}{\begin{equation}}
\newcommand{\ee}{\end{equation}}

 \makeatletter
 \g@addto@macro\normalsize{%
   \setlength\abovedisplayskip{15pt}
   \setlength\belowdisplayskip{15pt}
   \setlength\abovedisplayshortskip{15pt}
   \setlength\belowdisplayshortskip{15pt}
 }
 \makeatother
\begin{document}
\title{Coupled Electron-Ion Monte Carlo simulation of hydrogen molecular crystals}
\author{Giovanni Rillo$^1$, Miguel A. Morales$^2$,  David M. Ceperley$^4$, Carlo Pierleoni$^{5,6,*}$ }

\affiliation{
$^1$Department of Physics, Sapienza University of Rome, Italy\\
$^2$Physics Division, Lawrence Livermore National Laboratory, California USA\\
$^4$Department of Physics, University of Illinois Urbana-Champaign, Illinois, USA\\ 
$^5$ Department of Physical and Chemical Sciences, University of L'Aquila, Italy\\
$^6$Maison de la Simulation, CEA Saclay, France\\
}

\date{\today}
\begin{abstract}
We performed simulations
for solid molecular hydrogen at high pressures (250GPa$\leq$P$\leq$500GPa) along two isotherms at T=200 K (phases III and VI) and at T=414 K (phase IV). 
At T=200K we  considered  likely candidates for phase III, the C2c and Cmca12 structures, while at T=414K in phase IV we studied the Pc48 structure. 
We employed both Coupled Electron-Ion Monte Carlo (CEIMC) and Path Integral Molecular Dynamics (PIMD) based on Density Functional Theory (DFT) using the vdW-DF approximation.
The comparison between the two methods allows us to address the question of the accuracy of the xc approximation of DFT for thermal and quantum protons without recurring to perturbation theories. In general, we find that atomic and molecular fluctuations in PIMD are larger than in CEIMC which suggests that the potential energy surface from vdW-DF is less structured than the one from Quantum Monte Carlo. 
We find qualitatively different behaviors for systems prepared in the C2c structure for increasing pressure. Within PIMD the C2c structure is dynamically partially stable for P$\leq$250GPa only: it retains the symmetry of the molecular centers but not the molecular orientation; at intermediate pressures it develops layered structures like Pbcn or Ibam and transforms to the metallic Cmca-4 structure at P$\geq$450GPa. Instead, within CEIMC, the C2c structure is found to be dynamically stable at least up to 450GPa; at increasing pressure the molecular bond length increases and the nuclear correlation decreases. For the other two structures the two methods are in qualitative agreement although quantitative differences remain. We discuss various structural properties  and the electrical conductivity. We find these structures become conducting around 350GPa but the metallic Drude-like behavior is reached only at around 500GPa, consistent with recent experimental claims.
\end{abstract}

\maketitle

\section{Introduction}
Hydrogen is the first element of the periodic table and, as such,  is often
regarded as the simplest one. However, hydrogen shows a complex behavior in its condensed phases as
the density increases.
Speculations about the existence of  a low temperature metallic solid state  at 25 GPa started with 
Wigner and Huntington\cite{Wigner1935}; later calculations suggested
that this state could  become a high-temperature
superconductor\cite{Ashcroft1968}.
When experiments achieved the predicted transition pressure, they did not
find a metallic state but a rich phase diagram
with several different molecular structures\cite{Mao1994,McMahon2012a,Howie2015}.
The quest for metallic hydrogen at low temperature is still 
an on-going activity with different experimental methods providing conflicting results\cite{Zaghoo2016,Ohta2015,Knudson2015,Eremets2016,Dias2017}.

Performing experiments at  high pressures is complicated and the information obtained
is partial. At low temperature in the crystalline phase, the boundaries among the different solid phases are located by 
changes in the vibrational spectra, but most of  their 
structural properties remain elusive.
At least four different phases have been identified: the low pressure normal phase I, the broken-symmetry phases II and III\cite{Mao1994} and 
the mixed phase IV beyond 250 GPa and above 250 K\cite{Howie2012b,Howie2015}. Phase IV was later shown to be at most semimetallic\cite{Zha2012,Zha2013,Loubeyre2013}. Eremets et al. \cite{Eremets2016} studied hydrogen at pressures up to 380 GPa and T<200 K with Raman scattering. For P>360 GPa they find that the intensity of the low frequency Raman spectra vanish when cooling the system below 200 K; 
at the same time, a strong drop in resistance is observed in the same thermodynamic conditions
(P>360 GPa and T<200 K). They thus draw a vertical transition line in the P-T plane, introducing a  new conducting phase VI  for pressures higher than P=360 GPa.
Dalladay-Simpson et al. \cite{Dalladay-Simpson2016} investigate the system at $T\ge300$ K. 
They propose a new phase (V) for P>325 GPa, based  on arguments similar for phase transitions at lower pressures: 
changes in the low frequency peaks and in the slope of the pressure dependence of the vibron, with broadening and  weakening
of the vibrational peak itself. The Raman intensity, in general, decreases: this, coupled to the weak vibronic signal,  is interpreted as quasi-atomic state, a precursor of a fully non-molecular, metallic system. 
Dias et al.\cite{Dias2016} probe the system at low temperature (T<200 K), as in ref. \cite{Eremets2016}, using infrared radiation. Above 355 GPa, the IR vibron disappears,
and two new peaks close to 3000 $cm^{-1}$ appear; they come up with a vertical transition line, similar to the one proposed in ref. \cite{Eremets2016}. At variance with ref. \cite{Eremets2016} though,
no evidence of metallicity is found.
More recently, Dias and Silvera\cite{Dias2017} reported the observation of the Wigner-Huntington transition at 495 GPa and 5 K; this result has been much debated in the literature \cite{Goncharov2017,Liu2017,Loubeyre2017,Silvera2017}.

\begin{figure}[h]
\includegraphics[width=0.45\textwidth]{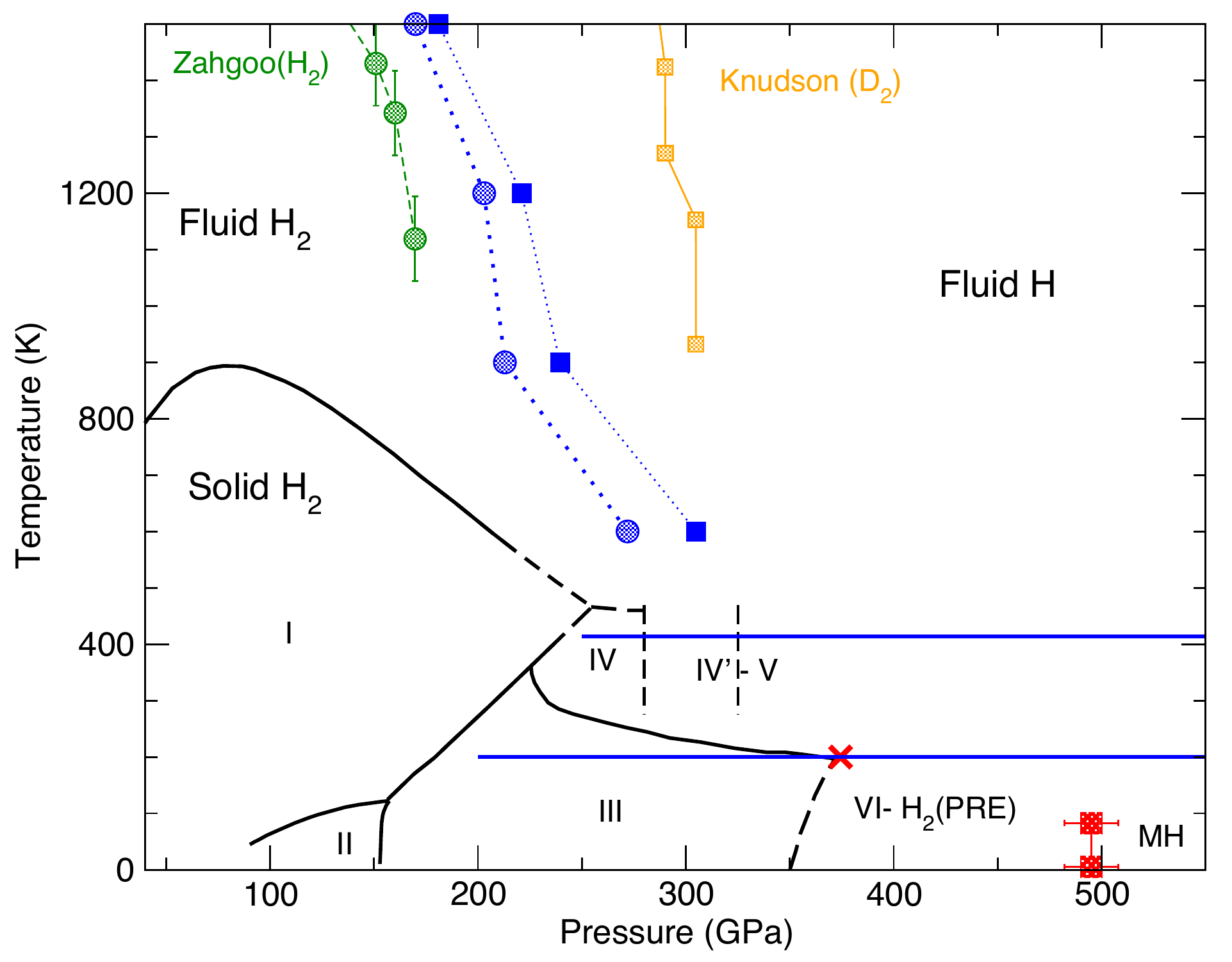}
  \caption{\small Phase diagram of hydrogen. The red cross indicates the condition at which semimetallic behavior has been observed \cite{Eremets2016}, while the red squares are the condition where Wigner-Huntington transition has been reported \cite{Dias2017}. The lines in the fluid phase reports recent claims of the observation of the liquid-liquid phase transition by different methods: static compression (green circles) \cite{Zaghoo2016}, dynamic compression for deuterium (orange squares) \cite{Knudson2015}, CEIMC simulations (blue circles hydrogen, blue squares deuterium)\cite{Pierleoni2016}. The blue horizontal lines indicate the isotherms  investigated in the present work.} 
\label{fig:phasediagram}
\end{figure}

The emerging phase diagram for solid hydrogen is shown in figure \ref{fig:phasediagram} together with recent determinations of the liquid-liquid transition line from different methods\cite{Zaghoo2016,Ohta2015,Knudson2015,Pierleoni2016}. 
Determination of the structure of phase III, which
is relatively well established in the solid phase diagram, is still an object of debate. 


{\it Ab initio} simulations can be a valuable tool to complement and interpret experimental data.  
Exploiting ab-initio random structure search (AIRSS) method within Density Functional Theory (DFT) theory and the PBE XC approximation,
Pickard et al.\cite{Pickard2007,Pickard2012}  found  several candidate structures for the various solid phases.
Although PBE generally predicts accurate geometries, the needed accuracy of the Potential Energy Surface (PES) away from the minima, and therefore the relative dynamical stabilities of the competing structures, is very high since the free energy differences between competing structures, and probably the amplitude of free energy barriers, is on the scale of the $\approx~$1  meV/atom \cite{Azadi2013b}. In addition, light nuclei display significant quantum effects;  perturbative treatment  of zero point effects is not necessarily adequate. 
These considerations motivated different simulation approaches. On the one side, Molecular Dynamics simulations at finite temperature were performed based on DFT to explore the dynamical stability of the proposed candidate structures\cite{Morales2013,Li2013,Liu2013,Liu2014,Chen2014}. These calculations could investigate both the influence of the exchange-correlation approximation and the nuclear quantum effects on the dynamical stability of the considered structures. On the other hand, static lattice energies were computed by the more reliable Quantum Monte Carlo methods\cite{Azadi2014,McMinis2015b,Drummond2015,Azadi2017}. However in these studies the treatment of finite temperature and nuclear quantum effects still relied on perturbation theory and the PBE-DFT functional.

The use of different exchange correlation functionals in conjunction with nuclear quantum effects for hydrogen was 
examined  by Morales et al. \cite{Morales2013liquid} in the liquid phase and then extended to solid hydrogen in ref. \cite{Morales2013}, where path integral molecular dynamics (PIMD) was employed along with the PBE, HSE and vdW-DF2 approximations to study structural properties of different crystalline
structures (C2c, Cmca12, Pbcn) at T=200 K. Important nuclear quantum effects were detected in structural properties, such as in the pair distribution functions. The choice of the $E_{xc}$ approximation was evident:  using vdW-DF2  instead of PBE results  in more pronounced molecular features, the opposite effect of the quantum nuclei.
Clay et al.\cite{Clay2014} have used QMC to benchmark a number of exchange correlation functionals for high pressure hydrogen and found that the vdW-DF exchange-correlation functional\cite{Dion2004} is the globally most accurate in this regime of density.

The aim of the present work is to perform simulations where  the $E_{xc}$ approximation, and the harmonic approximation, employed in static calculations, are replaced by controlled approximations, whose accuracy can be evaluated.
To obtain this result, we employ the Coupled Electron Ion Monte Carlo (CEIMC) method\cite{Morales2010,liberatore}. This method combines the Path Integral formalism to treat the nuclear
degrees of freedom and the Variational Monte
Carlo (VMC) method to accurately compute electronic energies within the Born-Oppenheimer approximation; it can perform finite temperature simulations without the uncertainty of 
DFT calculations. We apply the method  to the low temperature, solid phase, running 
finite temperature simulations of different candidate structures for phase III and IV. 
At the same time, we performed other simulations where  electronic energies  were computed through DFT, using the vdW-DF exchange correlation functional, and nuclear quantum effects were treated using Path Integral Molecular Dynamics (PIMD) with the Generalized Langevin Ensemble Dynamical Method \cite{Ceriotti2011} (GLE).  
 In particular, we focus on the T=200 K isotherm for pressures P>200 GPa: the recent claim of the discovery of a new solid phase which may be semi-metallic \cite{Eremets2016} makes
this region of the phase diagram a natural choice for our computations at finite temperature.
Another interesting region  involves the  boundaries of phase IV at higher temperatures ($\approx 400$~K),
close to the melting line.
Our PIMD data complete and extend the work by Morales et al.\cite{Morales2013}.

The simulation protocol adopted for PIMD and CEIMC simulations is described in section \ref{sec:methods};  our
results are presented in section \ref{sec:results} and discussed in section \ref{sec:discussion}, where they are compared to 
previous works. Final conclusions are drawn in section \ref{sec:discussion}.

\section{Methods}
\label{sec:methods}
We performed  simulations of solid hydrogen at finite temperature along two isotherms at T=200 K and T=414 K.
At T=200 K, for pressures between 250-500 GPa, we considered  C2c, Cmca12 and Cmca4 structures using
PIMD and  C2c and Cmca12 structures for CEIMC ( Cmca4 is excluded by static QMC calculation\cite{Azadi2014,McMinis2015b}).
At T=414 K we considered only Pc48  in a range of 250-350 GPa, since both DFT and QMC found this structure stable at higher temperatures \cite{Pickard2012,Pickard2012erratum,Drummond2015}.
The supercells used in the simulations contain 96 hydrogen atoms, a size that allows simulation boxes containing four distinct layers for the structures mentioned above.

Both PIMD and CEIMC were used to perform simulations at constant temperature T and volume V: 
consequently, each  thermodynamic point is identified 
by the  temperature and the parameter $r_s$, defined as $\frac{4}{3}\pi r_s^3a_0^3=\frac{V}{N}$, where $a_0$ is the Bohr radius.  
For each considered pressure,
the above structures were optimized at constant pressure using the vdW-DF approximation with the algorithms for geometry optimization implemented in Quantum Espresso\cite{Giannozzi2009a};
the resulting supercells were then used as a starting point for the PIMD and CEIMC simulations.
In both methods the Born-Oppenheimer approximation was used to separate the electronic from the protonic degrees of freedom.

%
%
%
%

\subsection{PIMD}  
In PIMD the Born-Oppenheimer energy surface was approximated by the vdW-DF functional\cite{Dion2004}, this functional is the most consistent  with Quantum Monte Carlo calculations\cite{Clay2014}. The equilibrium nuclear quantum effects are treated by the Imaginary time-path integral formalism where each proton is expanded into a path consisting of $P$ ``beads''.  Straightforward Molecular Dynamics of this description runs into several problems. First, the dynamics do not correspond to physical quantum dynamics, only the equilibrium properties are rigorously given by the simulation. Secondly, the resulting dynamics of the path integral degrees of freedom is not ergodic. It can be made ergodic by using Langevin dynamics where the system is coupled to fictitious heat bath. Thirdly, the computer time is proportional to the number of beads, since a full DFT calculation needs to be performed for each bead. The number of beads can be reduced using a Generalized Langevin Dynamics (PI+GLE). In addition the  ``colored noise''  makes the bead dynamics ergodic\cite{Ceriotti2011,Ceriotti2016}.

PIMD simulations were performed using a modified version of VASP
\cite{Kresse1996}, which incorporated the PI+GLE method  to generate the suitable NVT ensemble.   
16 beads were used at T=200 K, with a Trotter time step $\tau=0.0003125$ $K^{-1}$, the same value that  was used in ref. \cite{Morales2013}; 8 beads at T=414 K, with $\tau\approx0.0003019$ $K^{-1}$. The convergence in Trotter number was tested in ref. \cite{Morales2013}. A PAW pseudopotential\cite{PAW} was employed, with an energy cutoff of 350 eV and a 2x2x2 Monkhorst-Pack grid of $\vec{k}$ points. A time-step of 0.2 fs was used for the sampling dynamics. After an equilibration period of $\sim$0.25 ps, statistics were gathered for simulation times of $\sim$1.25--1.75 ps, corresponding to 6500--9000 time steps. This was found to be long enough to obtain well converged thermodynamic properties across the entire pressure and temperature range studied. Other details are given in ref. \cite{Morales2013}.

\subsection{CEIMC}
In Coupled-Electron Ion Monte Carlo (CEIMC), the Born-Oppenheimer (BO) energy is calculated using Variational Quantum Monte Carlo.
Then imaginary time path integrals are used to describe quantum nuclei.  Both electronic and protonic degrees of freedom are sampled using specialized Monte Carlo techniques and the ``penalty method'' is used to ensure that the resulting protonic distribution is unbiased with respect to the noise in the electronic Monte Carlo.

An accurate trial wavefunction is needed to compute electronic energies  within CEIMC. We use a Slater-Jastrow wavefunction with a small number of  
variational parameters; a  detailed description  is given in refs. \cite{Pierleoni2006,Holzmann2003,Pierleoni2008,Morales2014entropy}. 
Twist average boundary conditions (TABC)  using a grid of 4x4x4 twists were employed to reduce finite size effects. The  single particle orbitals were obtained 
through  a PBE-DFT self consistent calculation, with $E_{cut}=540$ eV with the same 4x4x4  grid of $\vec{k}$ points. 
Wave function optimization was performed on the initial protonic configuration (i.e. the perfect crystal) and for a single twist only. We minimize a linear combination of the electronic ground state energy and  electronic variance by a correlated sampling procedure.
Previous calculations \cite{Pierleoni2016} have shown that the accuracy of this procedure on the electronic energy is $\sim$ 0.02mH/atom. Higher accuracy can be reached by using Reptation QMC \cite{Pierleoni2016}.

We have developed specialized procedures to converge the protonic path integral with respect to the number of beads. Firstly within CEIMC, the computer time depends weakly on the number of beads because, in general, Monte Carlo integration is not very sensitive to the number of variables that are integrated over. Each bead becomes a separate random walk which can be run in parallel, and results in reducing the variance on the BO energy $E_{BO}$. The slow convergence in bead number comes when the $E_{BO}$ energy surface is highly non-linear with respect to the proton coordinates. We approximate the BO surface with a sum of pairwise potentials between protons, e.g. $V_{pair}=\sum_{i<j}  v(r_{ij})$ and calculate the pair density matrix corresponding to $v(r)$ \cite{Ceperley1995}. If the short distance behavior of the potential is a reasonable representation of the strong non-linearities at short distances between protons screened by the electrons, by subtracting $V_{pair}$ from the electronic BO surface, $E_{BO}-V_{pair}$ becomes much smoother 
and can be accounted for at the primitive approximation level\cite{Pierleoni2006,Ceperley1995}. 
We obtain the potential $v$ between pseudo-nuclei by a Boltzmann Inversion procedure matching the pair correlation functions from a CEIMC run for classical protons at fixed density and temperature in the liquid phase. Transferability of this potential to other thermodynamic conditions is not an issue here since we just require that the short distance behavior be reasonably well represented. We use the same effective potential employed across the liquid-liquid phase transition of ref. \cite{Pierleoni2016}. Since we are investigating similar density ranges here, we expect a similar convergence of the results with the number of imaginary time slices. Here we used 32 beads at T=200 K, with an imaginary time timestep $\tau=1.5625 \times 10^{-4}$ $K^{-1}$; and 14 beads at T=414K with $\tau= 1.7253 \times 10^{-4}$ $K^{-1}$. Both values are smaller than $\tau=2.083\times 10^{-4}$ $K^{-1}$ of ref. \cite{Pierleoni2016}.

Protonic path space is explored through a two--level collective--move Metropolis scheme. Collective moves are necessary since the many-electron wave function must be recomputed entirely for every individual proton move. For quantum nuclei we sample within the normal mode coordinates of the free particle path in order to decouple the amplitude of the centroid move from the internal modes\cite{Tuckerman2010}. Proton collective moves are performed with a reasonably high level of final acceptance by a Smart Monte Carlo scheme using forces \cite{Allen1987}. There is a great flexibility in choosing the pre-rejection potential, the only requirement is that the acceptance ratio at the first level be large enough not to bias the final sampling. We use the pairwise potential mentioned above and its forces. A better choice, which provides a faster final equilibration of the nuclear degrees of freedom, is to use DFT energies and forces to pre-screen the collective moves\cite{Pierleoni2016}. In the present calculation, where the temperature is low, the first acceptance ratio is always rather large ($\geq 85\%$) while the final acceptance, determined by the large noise level, is around $30\%$. We find that the overhead of performing the DFT calculation even for moves rejected at the first step is negligible. With this scheme we can sample systems containing 100 protons and paths of 32 beads. 

It is interesting to compare how quickly the two simulation methods move through phase space. To make this comparison we have computed the diffusion time of the single proton away from its average position. We found that  CEIMC is roughly 20 times slower than the PIMD (in units of global move attempts).  To compensate we run CEIMC for  roughly 10 times more global moves than the PIMD simulations. For both algorithms we find stationary states obtained after a warm up period as we discuss below.

\subsection{Optical calculations}
One longstanding issue is whether solid molecular hydrogen undergoes metallization when pressure
is increased or whether metallization requires a transition to an atomic phase. We can see how
C2c and Cmca12 behave in this regard by studying their static electrical conductivity.
Within DFT, electrical conductivities
can be computed for a fixed nuclear configuration using the Kubo-Greenwood formula\cite{Kubo1957a,Greenwood1958}.
In our case, we select 16 protonic configurations sampled during the PIMD simulations and compute an average
conductivity. 
In the QMC approach, rigorous calculations of the conductivity
involve many-body excited states: there is no affordable and easy scheme to compute conductivities using
Quantum Monte Carlo methods, see e.g. ref. \cite{Lin2009}.
Therefore we use the  proton configurations sampled by CEIMC
to perform the Kubo--Greenwood calculation with DFT. 
The computation of  the  electrical conductivity  was performed using  
the HSE hybrid functional \cite{Heyd2003} which, while being computationally expensive, is known to  reproduce experimental  band gaps for many semiconductors \cite{Heyd2005} and was already successfully used in ref. \cite{Morales2013liquid}. The PAW pseudopotential was the same one that was employed for the PIMD runs.
We used an energy cutoff of 500 eV and  a 4x4x4 Monkhorst-Pack grid of $\vec{k}$ points;  for each $\vec{k}$ point, 64 bands were employed with a smearing
parameter of 0.3 eV.  The static DC value of the conductivity, $\sigma_0$ was obtained by extrapolating a polynomial fit of $\sigma (\omega)$  to $\omega = 0$.  

\section{Results}
 \label{sec:results}
 To analyze molecular crystals, we are interested in studying the
stability of the structure and  the
orientational order of the molecules. 
Following ref. \cite{Morales2013}, we introduce the orientational order
parameter (OOP):
\begin{align}
\hat{O} = \langle \Big [ \frac{1}{N} \sum_{i=1}^N P_2 (\hat{\Omega}_i \cdot \hat{e}_i) \Big ]^2 \rangle
\end{align}
where $P_2$ is a Legendre polynomial,
$\hat{\Omega}_i$ is the orientation of molecule $i$ during the
simulation, $\hat{e}_i$ is its initial orientation in the static
lattice  and N is the number of molecules. The order parameter equals 1 if the molecules
stay aligned with respect to their initial orientation while it equals 0 if they assume a random orientation. 

We also compute the molecular Lindemann ratio ($\hat{L}$)
(MLR):
\begin{align}
\hat{L}= \frac{\sqrt{\Delta^2}}{d},\qquad  \Delta^2=\langle \frac{1}{N}\sum_{i=1}^N (\mathbf{r}_i-\mathbf{r}_{i0})^2\rangle
\end{align}
where $d$ is the nearest neighbor molecular distance in the static lattice,
$\mathbf{r}_i$ is the position of the center of mass of molecule $i$
during the simulation and $\mathbf{r}_{i0}$ the position of its center
of mass in the static lattice. Note that for simple monoatomic lattices, the Lindemann ratio at melting reaches $\sim 0.15$ and $\sim 0.30$ for classical and quantum systems respectively.

Since we are interested in mixed structures, these
observables can be computed for every single layer in the simulation box to better
characterize the structures. In the same way, we can define a
layer-by-layer pair correlation function $g_{pp}(r)$, when only atoms belonging to
the same layer are taken into account, as done in ref. \cite{Chen2014}.
These layer-by-layer $g_{pp}(r)$, being normalized as in the bulk case, do not go to 1 for large r; however,  they are significant for distances shorter than the interlayer distance, roughly $\lesssim$2-3 a.u..
\subsection{C2c, T=200 K}
\begin{figure*}
\includegraphics[width=0.8\textwidth]{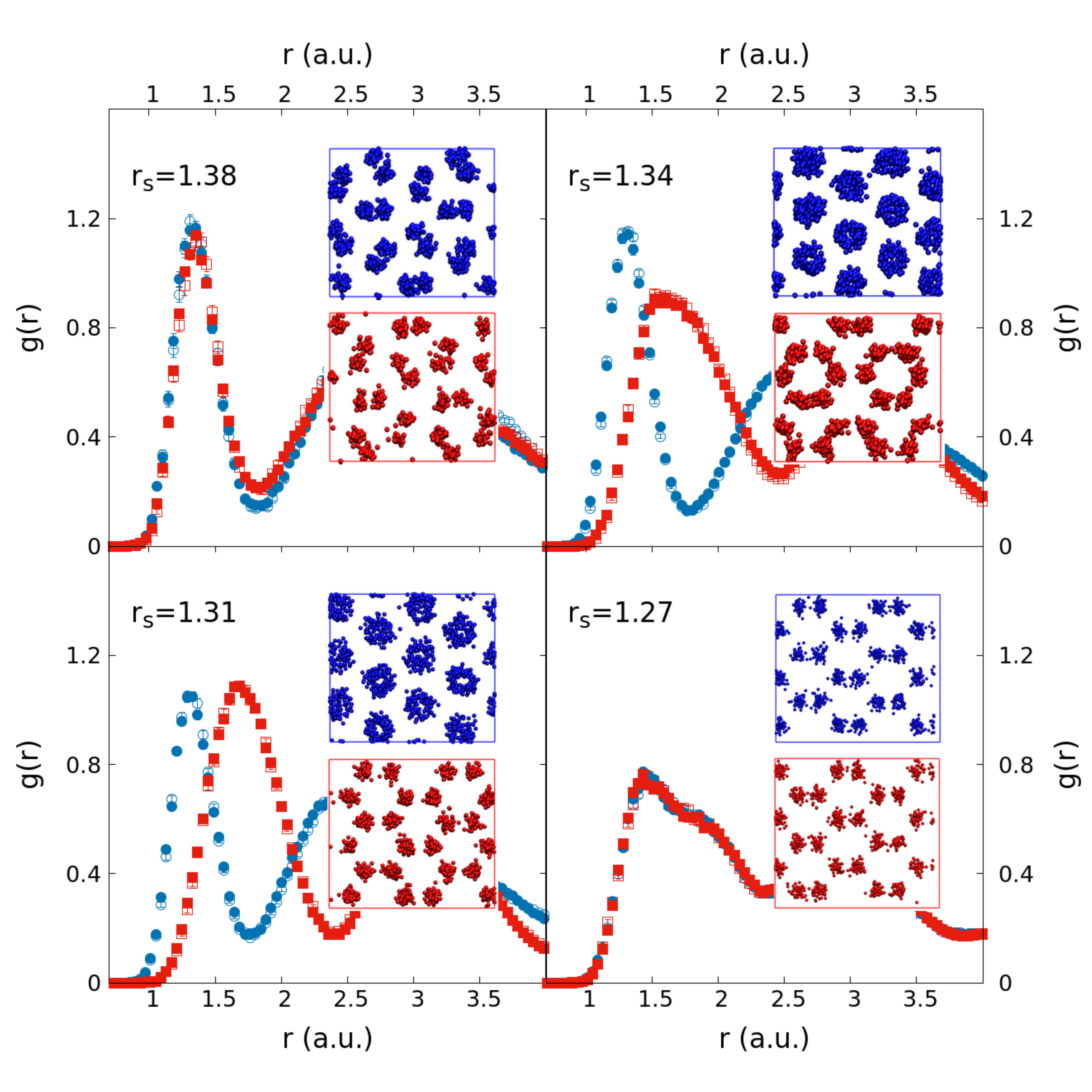}
  \caption{\small Layer-specific pair correlation functions  for  four
    densities from PIMD simulations starting from the C2c lattice.   
    Each simulation cell contains four distinct layers: their pair correlation
    functions are reported with different symbols (open and closed squares and circles). Layers behaving in the same
    way have the same color. Insets: snapshots of proton configurations from two
  adjacent layers.} 
\label{fig:pcfC2c}
\end{figure*}
We start our analysis  considering the PIMD simulations starting from the
C2c structure, which is  one of the main candidates for phase III. We detect different structural rearrangements at increasing density
using the
layer-by-layer pair correlation functions  as shown in
Fig. \ref{fig:pcfC2c}.
Note that there are four distinct planar layers in the simulation cell, all equivalent in the crystalline structure. At the lowest density considered ($r_s=1.42, P\simeq 200$GPa, not shown in Fig. \ref{fig:pcfC2c}), they all exhibit the same pair correlation function. With increasing
pressure, a structure with alternating, non-equivalent layers emerges. At the highest pressure, the layers become  equivalent again.

A qualitative understanding is gleaned
by looking at  the equilibrium nuclear configurations: see insets in fig. \ref{fig:pcfC2c}.
At $r_s=1.38$, we have four nearly equivalent layers of well-defined molecules:
the molecular centers retain their initial C2c symmetry, although with large fluctuations, and the molecular axes fluctuate around their initial orientations but are found to undergo large collective motions as discussed later.
At $r_s=1.34$ we observe a mixed structure of alternating layers: ``blue'' layers of strong molecules ($r_{mol}\sim 1.4a_0$) with large in--plane rotational activity, alternated by ``red'' layer of weak molecules ($r_{mol}\sim 1.5-1.6a_0$) with in--plane ring structure and reduced libration activity.
Computing the average atomic positions in the red layers shows a slight departure from
the hexagonal symmetry.  
These features indicate a Pbcn structure already proposed for molecular hydrogen \cite{Pickard2007}.
At $r_s=1.31$ this mixed structure changes: ``blue'' layers remain stable with a reduced proton correlation at the same bond length. ``Red'' layers transform to graphene--like atomic layers (Ibam symmetry). Finally, at $r_s=1.27$,  the layers are again all equivalent and the system
has  Cmca4 symmetry.

To discuss the molecular nature of the hexagonal layers at $r_s=1.31$,  we computed
the electronic charge density,  assuming that the existence of a molecular bond is related to a build up  of electronic charge between two protons.
Plots of the charge density for a single protonic configuration suggest the existence
of both molecules and isolated atoms: during the simulation, the
atomic motion results in a fast bonding and rebonding activity among
different atoms. 

\begin{figure}[h]
\begin{center} 
\includegraphics[width=0.45\textwidth]{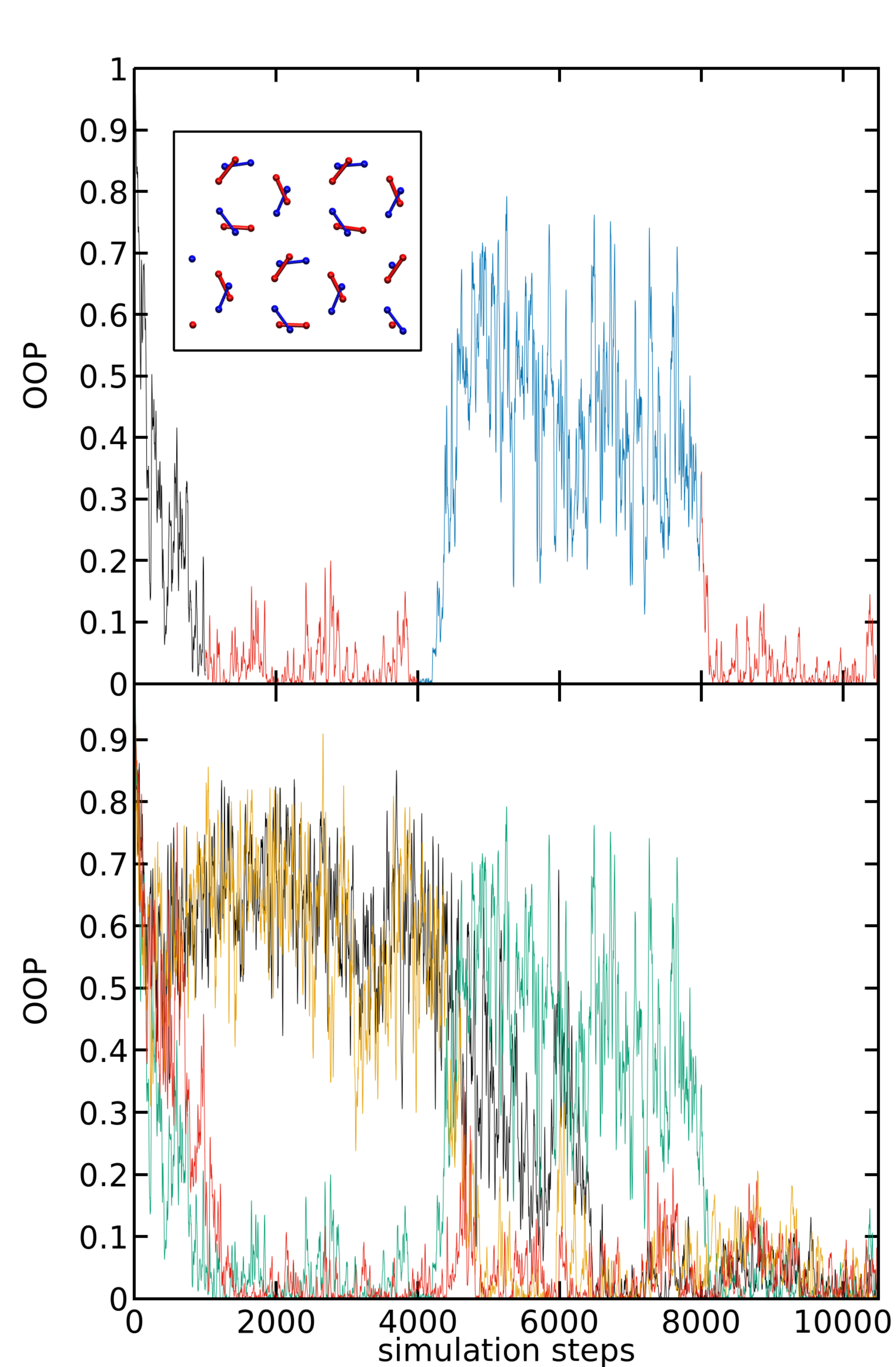}
\end{center}
\caption{\small Orientational order parameter (OOP) along the PIMD trajectory starting from
C2c symmetry at $r_s=1.38$ and T=200 K.  Upper panel: trace for a single layer. Segments of the trajectory which display a similar value of the OOP  are colored in red (OOP$\approx0$) or blue (OOP$\approx0.5$). The insert shows the positions of the atoms in the layer by averaging over the corresponding blue (red) portion of the trajectory.
Lower panel: trace of the orientational order parameter (OOP) for the four layers represented by different colors: layer 1-black, layer 2-green, layer 3-yellow, layer 4-red.}
 \label{fig:oopc2c}
 \end{figure}

An interesting phenomenon takes place at $r_s=1.38$ and $r_s=1.42$, where  the system keeps its
original symmetry: the orientational order parameter (OOP), which is used to detect changes in the alignment of 
the molecules with respect to their initial orientation, displays  sudden ``jumps'' along the trajectory. This is shown 
in the top panel of fig.~\ref{fig:oopc2c}, where the evolution of the  OOP of a single layer at $r_s=1.38$ is pictured. 
At the beginning of the trajectory, all the molecules are aligned in their perfect crystal direction and the OOP is $1$; 
after $\approx$ 1000 time steps, the orientational parameter goes to zero, revealing a rearrangement of the molecular axes; 
after 3000 steps there is a sudden increase of the parameter around the value 0.5 which remains 
stationary for 4000 steps; finally, the OOP goes back to 0 with another sudden jump after roughly 8000 steps. We can separate the trajectory into segments according
to the values assumed by  the OOP, and take the average atomic positions for each segment. The  ``jumps'' of the parameter correspond
to a collective rotation of the molecules; in particular, we can identify rings of three molecules (trimers) rotated by the same angle (inset in the top panel of fig.~\ref{fig:oopc2c}).
These events are rare on the time scale of our simulations, occurring once or twice per run. It is interesting to note that
these collective rotations can occur independently for the different layers, as reported in the bottom panel of fig.~\ref{fig:oopc2c}: the orientational parameters of layer 1 (black) and 3 (yellow)
follow the same trend, decaying to zero after 5000 time steps; layer 4 (red) goes to zero after the first 1000 steps; layer 2 (green) was described above. Layers with significantly different OOP can coexist at the same time, showing that the system as a whole can go through configurations where the C2c symmetry may not be preserved. Note that the layer by layer pair correlation functions are invariant under these rotations, since they preserve the distances between atoms in the same layer.

Since PIMD calculations are less computationally demanding, we ran more
simulations starting from the Pbcn and the Cmca4 structure at
different densities and observe the final symmetry of the system. These results are summarized in table \ref{tab:table1}. The stability ranges of the different phases depend on the initial
 structure. This is expected, since our simulations are
 performed at constant volume and   the system is constrained by the
 initial geometry of the supercell. Nevertheless, we can find a common
 chain of transitions as the volume decreases: C2c $\rightarrow$ Pbcn $\rightarrow$ Ibam
 $\rightarrow$ Cmca4. 
\begin{table*}
\begin{center}
 \caption{Dependance of the final structure on
  the initial structure as a function of density using PIMD. }
\label{tab:table1}
  \begin{tabular}{c|ccccccc}
\toprule
\diagbox{Starting \\
Structure}{$r_s$} & 1.42 & 1.38 & 1.34 & 1.31 & 1.29 & 1.27 & 1.25 \\
\midrule
C2c & C2c & C2c      & Pbcn   & Ibam  & Ibam  & Cmca4    & Cmca4      \\
Cmca4 & C2c & Pbcn   &  Ibam    &   Ibam  &  Cmca4   &   Cmca4  &     \\
Pbcn & C2c &  Pbcn    & Pbcn   & Ibam     &   Ibam   & Ibam    &     \\
Cmca12 & Cmca12 &  Cmca12    &  Cmca12   &  Cmca12   &  Cmca12   & Cmca12    &     \\
\bottomrule
\end{tabular}
\end{center}
\end{table*}
\begin{figure*}
  \centering
  \includegraphics[width=0.8\textwidth]{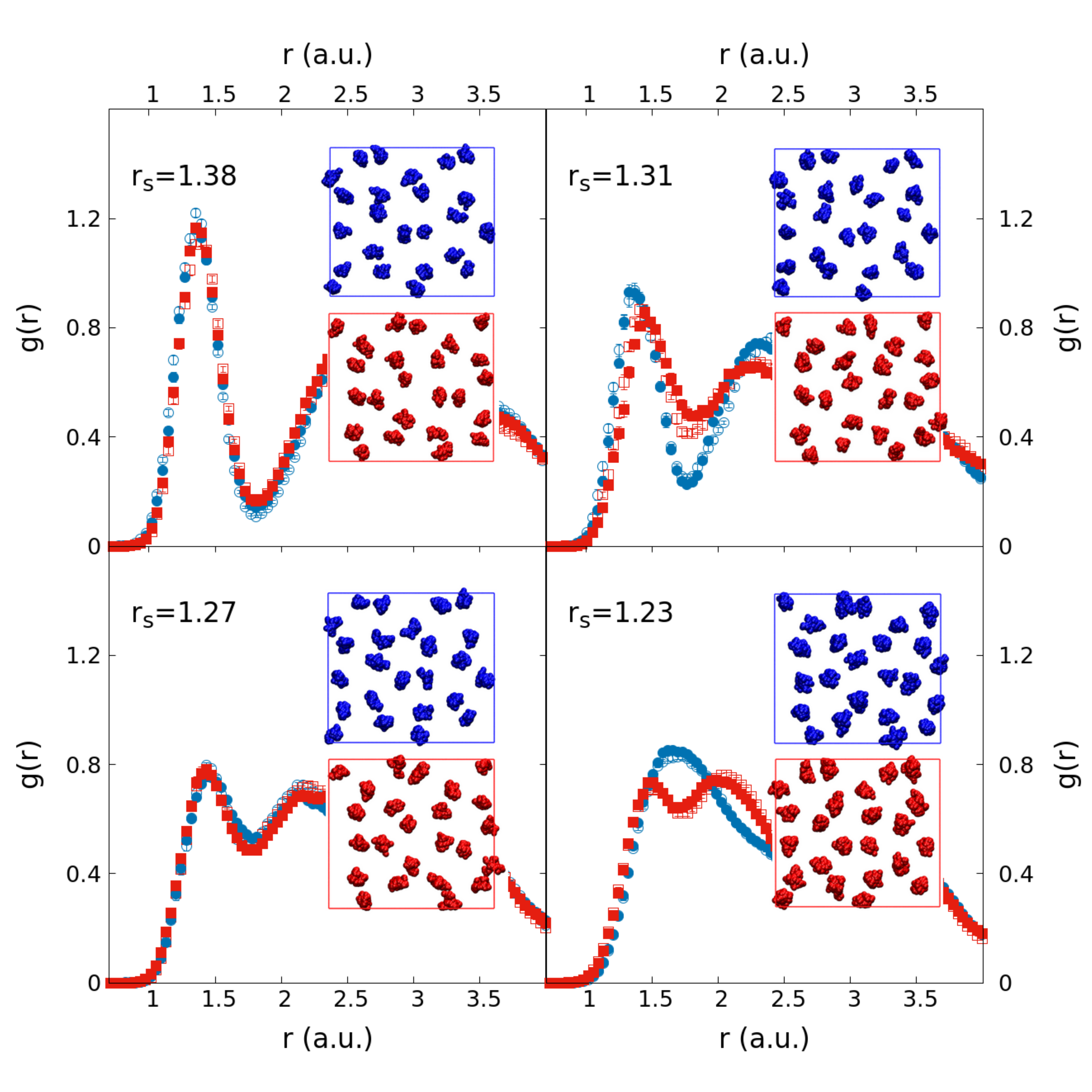}
  \caption{Layer-specific pair correlation functions from CEIMC simulations with C2c initial structure at four
    densities. Insets: snapshots of the proton configurations in two
    adjacent layers}
  \label{fig:CEIMCpcfC2c}
\end{figure*}

\begin{figure}
  \includegraphics[width=0.5\textwidth]{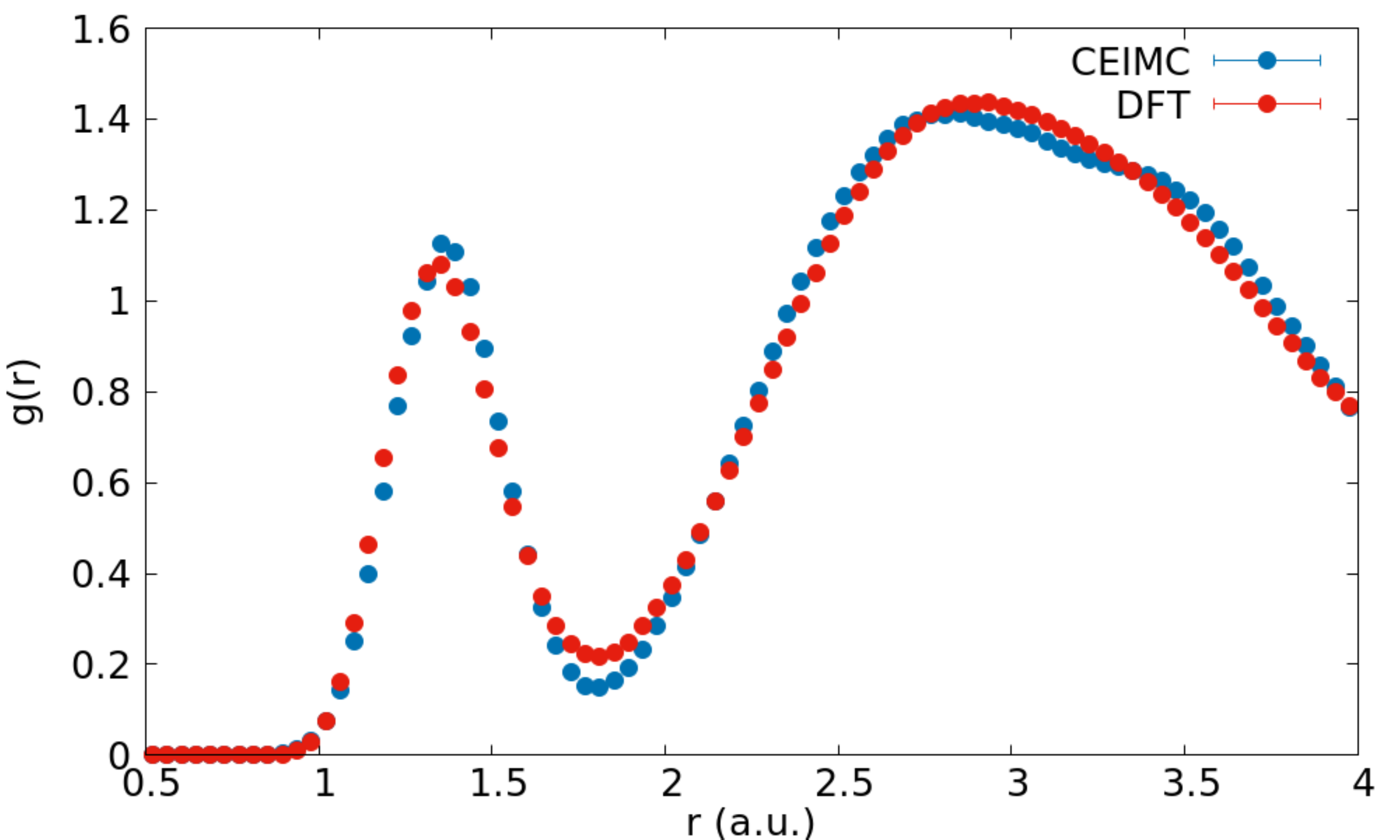}
\caption{Pair correlation function starting from the
  C2c structure at $r_s=1.38$. Comparison between CEIMC (blue) and PIMD (red).}
\label{fig:C2cdiff}
\end{figure}

We performed CEIMC runs starting from the C2c structure at $r_s=1.38,1.31,1.27,1.23$ ($200 < P  < 550$ GPa).
Values of average thermodynamics quantities like pressure, internal energy and enthalpy per atom are given in table \ref{tab:ceimc} and figure \ref{fig:CEIMCpcfC2c} reports results of the layer analysis.
\begin{table*}
\begin{center}
\caption{CEIMC data for the density dependence of pressure, energy and enthalpy for the structures considered. Statistical errors on the last digit are reported in brackets.}
 \label{tab:ceimc}
\begin{tabular}{|c|c|ccc|}
 \toprule
lattice& $r_s$& P(GPa)& $e_t$(h/at)& $h_t$(h/at)\\
\midrule
C2c& 1.380 &     242.8(1)  & -0.52688(4)   &-0.43605(6)   \\ 
 &1.314 &     337.4(4)  & -0.51179(5)  &-0.4029(1)   \\
 &1.266 &    426.1(4)  & -0.4987(1)  & -0.3755(2)  \\
& 1.230 &    517.9(3) & -0.48739(5)   & -0.35013(8)   \\
\bottomrule
Cmca12& 1.378 &     239.4(4) &  -0.52589(4) & -0.4367(1)   \\ 
& 1.312 &     331.0(5) &  -0.51093(6) &  -0.4044(2)   \\
& 1.265 &    426.1(4) &  -0.49805(5) &  -0.3753(1)    \\
 \bottomrule
  Pc48&   1.380 &  238.3(4) &    -0.52566(9)&  -0.4366(2) \\
   &  1.343 &  291.5(9) &  -0.5185(1)  &  -0.4179(7)\\
   &  1.313 & 335.7(6)  &  -0.5110(1)&  -0.4028(2)\\
\bottomrule
\end{tabular}
\end{center}
\end{table*}
\begin{figure*}
  \centering
  \includegraphics[width=0.8\textwidth]{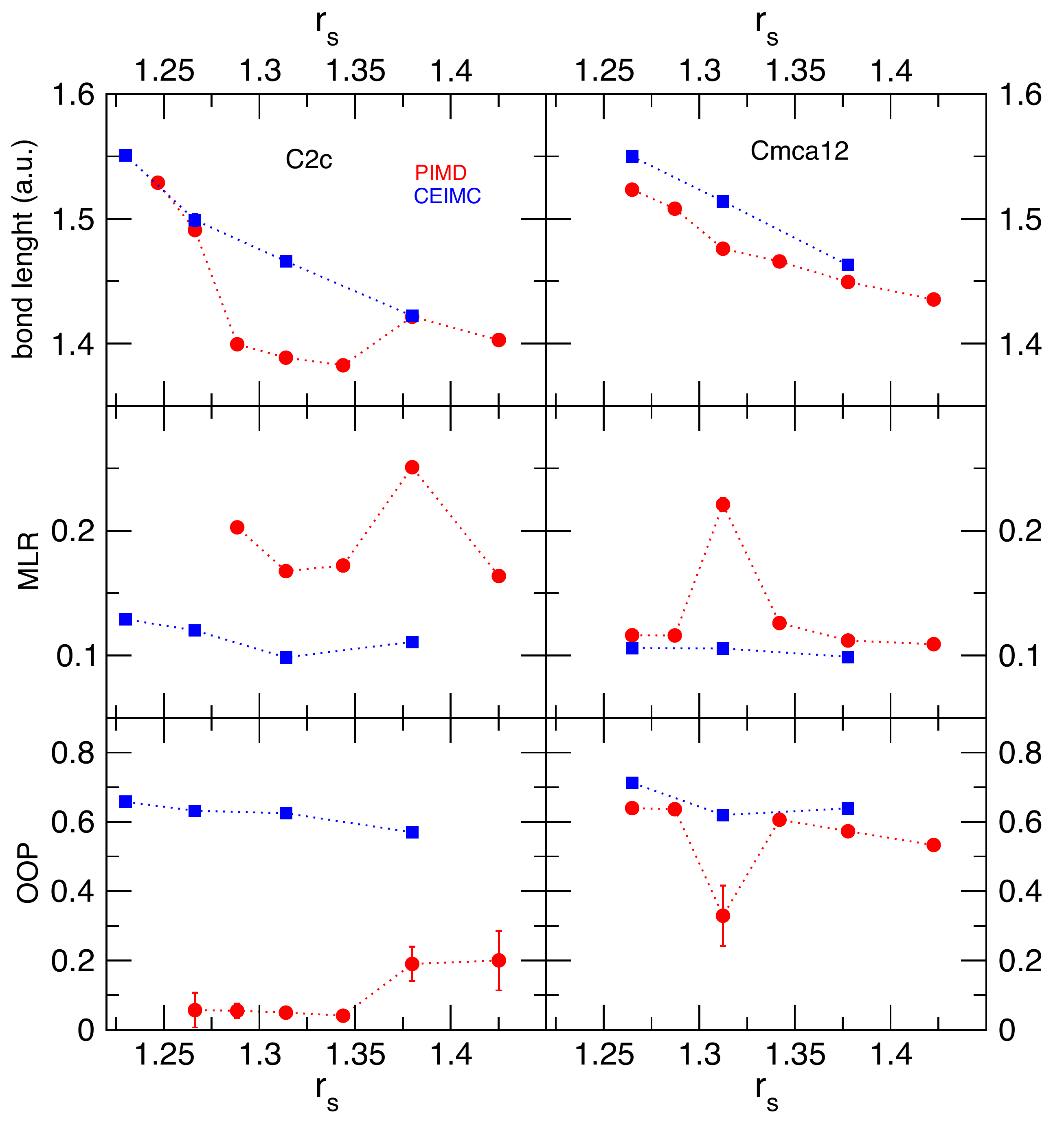}
  \caption{Molecular related structural properties for simulations starting
    from the C2c (left panels) and the Cmca12 (right panels) structures from both CEIMC (blue) and PIMD (red). 
    Top panels: average molecular bond length. 
    Middle panels: molecular Lindemann ratio (MLR).    
    Bottom panels: orientational order parameter (OOP).     
   Since for C2c, PIMD exhibits transitions to mixed structures with atomic planes, the reported molecular properties from PIMD correspond to the molecular layers only.}
  \label{fig:doppietta}
\end{figure*}
We observe that, except at the highest density, the main molecular peak around $\simeq 1.4 a_0$ is present in all layers, indicating that the system retains its molecular structure although with a decreasing amplitude. The same conclusion can be reached by visual inspection of proton configurations (see the insets in the first three panels of Fig. \ref{fig:pcfC2c} ). At $r_s=1.31$ alternating layers exhibit an alternating molecules, as evidenced by the amplitude of the first minimum ($r \simeq 1.75 a_0$), the ``blue'' layers having stronger molecular character than the ``red'' layers. 
While in the PIMD simulations the system adjusts to the increasing pressure by changing structure, in CEIMC the adjustment is primarily to decrease the proton correlation since the amplitude of the peaks is progressively decreasing. However at $r_s=1.27$ proton correlation in CEIMC is stronger than in PIMD.
In CEIMC, even at a higher density of $r_s=1.23$, an alternating molecular-non molecular layered structure appears with
a very weak molecular character in the ``red'' layer and a clearly non-molecular character in the ``blue'' layer (the first maximum is at $\approx 1.7a_0$).
The insets in fig. \ref{fig:CEIMCpcfC2c} show that clouds associated with individual
atoms can be spotted at all densities. This means that, when molecules exist, they have a reduced rotational activity, similar to what happens in the PIMD simulations at low density ($r_s=1.38$ and $r_s=1.42$).  During the CEIMC simulations, we do not observe the abrupt jumps in the orientational order parameter, which were present in the PIMD case. In the CEIMC simulations the OOP reaches stationary values consistent with the high plateau values found during the PIMD trajectories and corresponding to average equilibrium librations around the initial orientation.
Average values of the OOP from CEIMC are always around 0.6 as reported in figure \ref{fig:doppietta}. Note that the PIMD values shown in the same figure correspond to the molecular layers only. 
The absence of re-orientation or rotation in CEIMC is also reflected in the different molecular Lindemann's ratio (MLR): the MLR value from PIMD at $r_s=1.38$ and $r_s=1.42$ are larger than the CEIMC values, because the molecular centers move when the trimers rotate. On the other hand, the two pair distribution functions, compared in fig. \ref{fig:C2cdiff}, match each other rather well.  

A direct comparison between CEIMC and PIMD calculated properties at higher pressures is biased by the structural transitions observed within PIMD. 
When we observe mixed structures, the molecular properties are computed for the molecular layers only. 
The bond length is very sensitive to the changes of phase as seen in the upper panels of fig. \ref{fig:doppietta}. 
Indeed for the C2c structure, we can see a first discontinuity between $r_s=1.38$  and $r_s=1.34$ in the PIMD results, when the the first transition occurs: 
molecules in the mixed phases have a smaller bond length than in homogeneous CEIMC derived phase at the same density. 
A second discontinuity in the PIMD results is observed between $r_s=1.29$ and $r_s=1.27$, where Cmca4 becomes the stable phase in DFT with a considerably larger bond length.

With CEIMC, the system does not show tendency to undergo phase transition towards mixed phases but rather remains in the initial C2c structure. 
The effect of increasing pressure is twofold: the bond length increases with pressure, as seen in figure \ref{fig:doppietta}. 
At the same time the intermolecular distance decreases (the second peak of g(r) is getting closer to the first peak, see figure \ref{fig:CEIMCpcfC2c}) 
without changing much the molecular orientation (OOP) or the crystal symmetry. 
At $r_s=1.31$ we observe some tendency to express a two layer structure but the differences are minor in comparison to the PIMD results. 
Only at the extreme density of $r_s=1.23$  does the system go to a mixed structure with molecular layers alternating  with non-molecular layers. 
However this density corresponds to a pressure of $\approx 520$ GPa which is expected to be beyond the limit of stability of the molecular phase with respect to the atomic phase. 

The differences in the OOP and in the MLR between PIMD and CEIMC results suggest that the
energy barriers preventing molecular rotations and vibrations and displacements of the
molecular center is underestimated by PIMD for the densities at which the C2c structure is stable.

\subsection{Cmca12 at T=200 K}
\begin{figure}
\centering
  \includegraphics[width=0.45\textwidth]{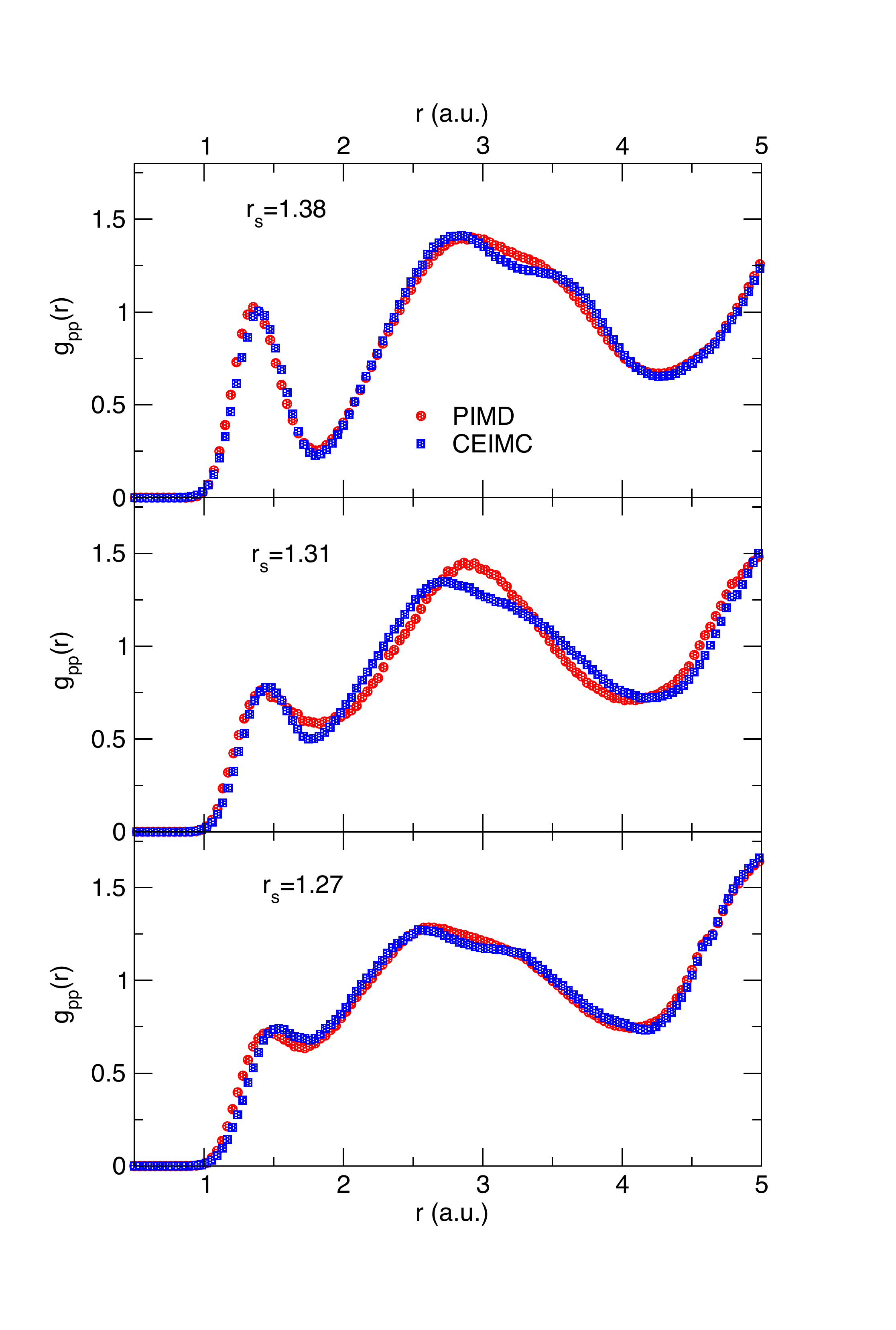}
\caption{PIMD and CEIMC pair correlation functions for systems starting from the
  Cmca12 structure at three densities.}
\label{fig:Cmca12diff}
\end{figure}

\begin{figure*}
\centering
  \includegraphics[width=0.8\textwidth]{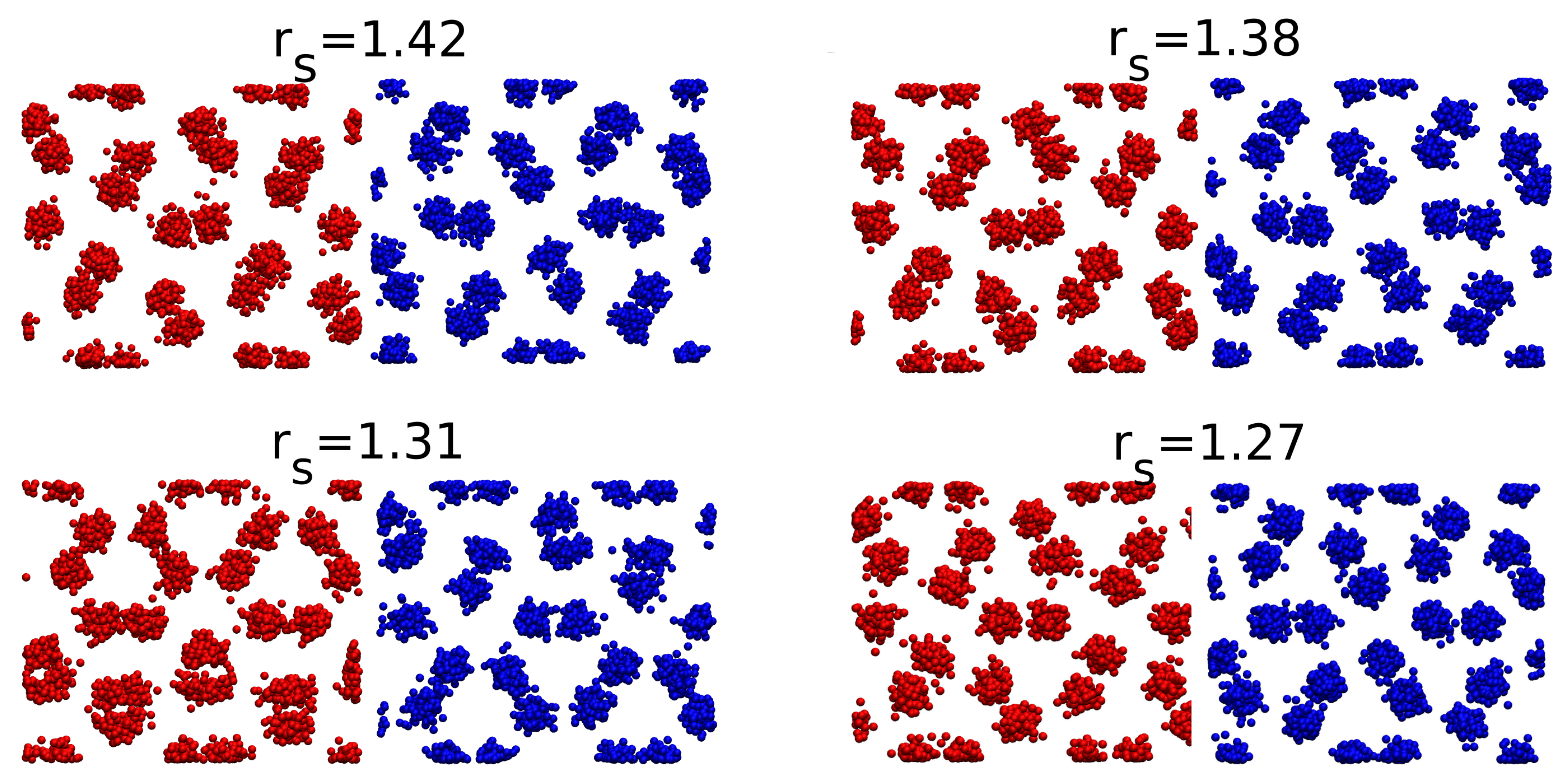}
\caption{Snapshops of proton configurations from PIMD starting from the
  Cmca12 structure in two different layers.}
\label{fig:Cmca12layers}
\end{figure*}

Systems starting from the Cmca12 structure do not display transitions
to structures with different layers with both simulation methods; in this case we can readily
compare PIMD and CEIMC results. In figure \ref{fig:Cmca12diff}  we compare the proton-proton pair correlation function at three densities. At 
$r_s=1.38$ and $r_s=1.27$ the two simulations are in good agreement although PIMD data exhibit a slightly shorter molecular bond (see also the left upper panel of figure \ref{fig:doppietta}). 
CEIMC data for the second maximum also exhibit a clear two-peak structure with a main maximum $\simeq 2.7 a_0$ and a shoulder at a larger distance, $\simeq 3.5 a_0$. 
PIMD data instead exhibit a single rather large second peak: the difference is probably due the slightly larger rotational and vibration activity of PIMD with respect to CEIMC data (see the left panels of figure \ref{fig:doppietta}). 
At $r_s=1.31$ the two pair distribution functions show a noticeable difference both in the first minimum and the second maximum. This discrepancy is 
the result of an incipient structural rearrangement, observed only in the PIMD trajectory. The rearrangement affects the MLR and the OOP values from PIMD shown in the left panels of fig. \ref{fig:doppietta}.
The rearrangement occurs after a long period of apparent stability of the Cmca12 structure, indicating that the Cmca12 structure within PIMD is dynamically much more stable than the C2c structure. 
Curiously, in this new structure, all of the layers remain equivalent but within each layer we have
both molecules forming rings and molecules rotating in--plane as shown in Fig. \ref{fig:Cmca12layers}.
We have not seen this instability at other densities but it is possible that it will occur after an even longer time.
The CEIMC trajectory does not show any sign of structural rearrangement, again suggesting that the QMC energy surface is more structured. 

Focusing on the densities where the Cmca12 symmetry persists, ($r_s=$1.38 and $r_s=$1.27)),
we see that MLR, OOP and bond length
between PIMD and CEIMC are comparable (fig. \ref{fig:doppietta}); in particular, the OOP is fairly high,
indicating  a reduced rotational activity, in contrast to what happens
in the PIMD simulations for C2c. The exception is, as reported above,
$r_s=1.31$ where some molecules rotate in the plane.
This comparison suggests that the
vdW-DFT functional reproduces well the properties of the Cmca12
symmetry while it artificially enhances the rotational activity of the
molecules in the C2c phase.

\subsection{Electronic properties}
\begin{figure*}
  \centering
  \includegraphics[width=0.8\textwidth]{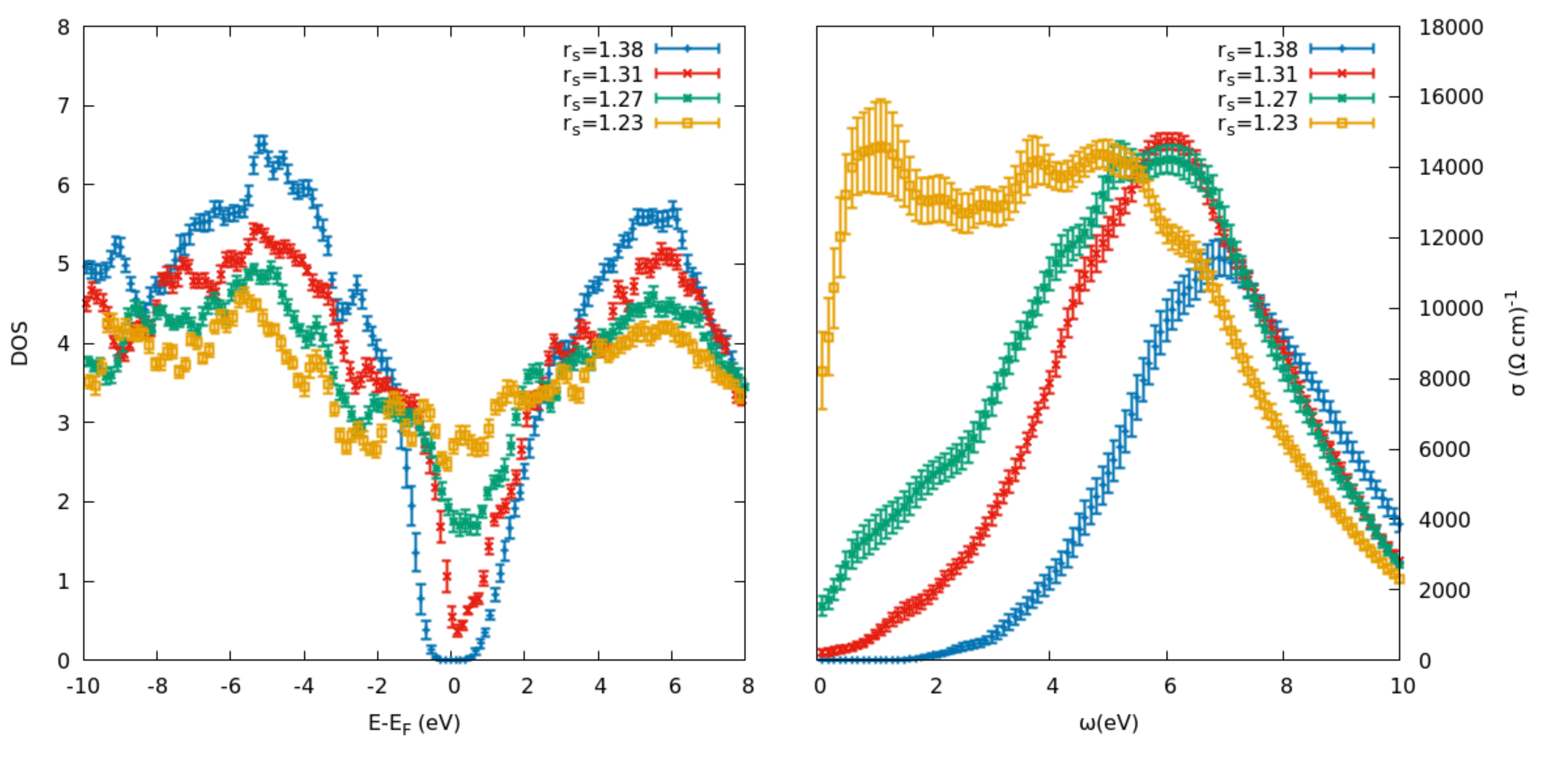}
  \caption{Left panel: electronic density of states averaged over 16 different proton configurations sampled during CEIMC runs
  starting from the C2c structure.  Right panel: the corresponding frequency dependent, longitudinal (in--plane) conductivity.}
  \label{fig:C2cdoscond}
\end{figure*}

In the right panel of fig. \ref{fig:C2cdoscond} we show the electronic density of states obtained by HSE-DFT, for 16 proton configurations sampled during the CEIMC runs in the C2c structure. In the left panel we report the longitudinal (in-layer) energy dependent conductivity. 
At $r_s=1.31$ the band gap closes and the system starts behaving as a bad metal; however, a dip remains in the density of states at the Fermi level, which results in a peculiar form of the frequency dependent conductivity and a rather small value of the DC conductivity. As the density increases, the dip at the Fermi level fills in, and the conductivity increases to reach a Drude-like behavior at $r_s=1.23$ with a consequent increase in the DC conductivity to a typical metallic value. This process occurs over a range of pressures of roughly 200GPa, pointing to the progressive character of the metallization process in these molecular crystals. A possible different mechanism would be an abrupt metallization driven by a structural phase transition to a monoatomic phase as proposed by ground state QMC calculations \cite{McMinis2015b,Azadi2014}, a mechanism that remains out of the reach of our simulations.

\begin{figure*}
  \includegraphics[width=0.7\textwidth]{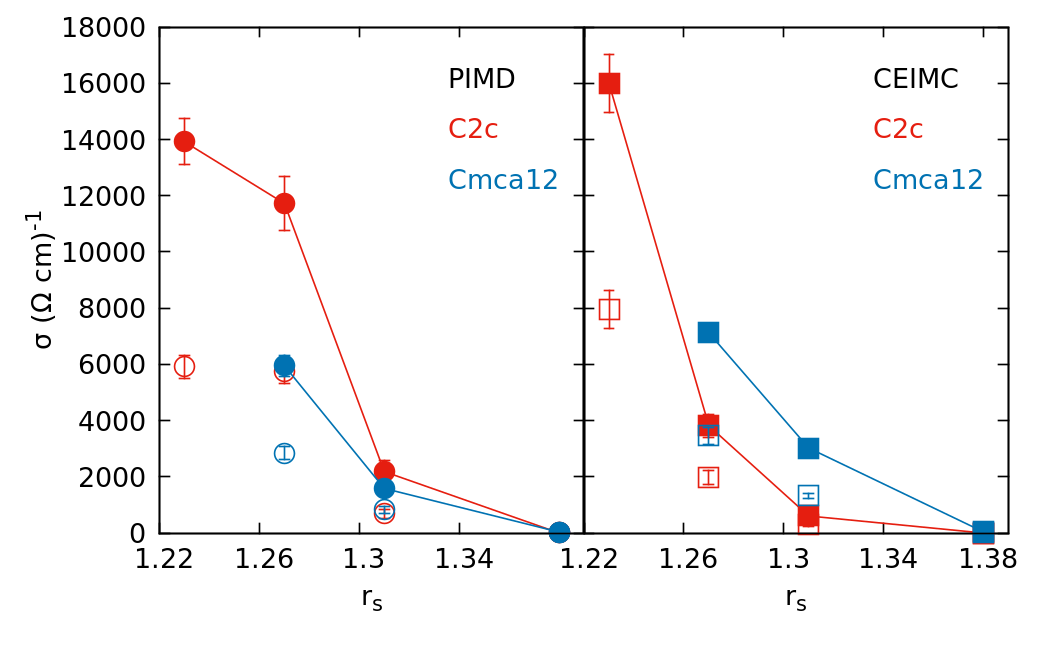}
  \caption{DC conductivity for systems in the C2c (red) and Cmca12 (blue) structures obtained from PIMD (left panel) and CEIMC (right panel) simulations: the longitudinal (parallel to the layers) conductivity is shown by filled symbols and lines, while the 
  transverse conductivities by empty symbols.  }
  \label{fig:cond1}\end{figure*}

Figure \ref{fig:cond1} reports the density dependence of the static conductivities for both C2c and Cmca12 structures. 
In the left panel we show the results obtained using nuclear configurations from the PIMD trajectories, while the right panel shows results for nuclear configuration from the CEIMC trajectories. 
Both longitudinal (in--plane) and transverse conductivities are computed, and we systematically find that the transverse conductivity is smaller than the corresponding in-plane one. 
Qualitatively, the behavior of the static conductivity is similar for the two structures, C2c and Cmca12, and for the two methods: the systems become metallic at $r_s \sim 1.31$, corresponding to a CEIMC pressures of P$\sim$ 335GPa (see fig. \ref{fig:cond1} and table \ref{tab:ceimc}).  
The relatively small values of the conductivity suggest that these structures are semi-metallic in this range of densities.
Within CEIMC, the conductivity of the Cmca12 structure is systematically higher than
in C2c; this is probably related to the larger bond lengths of the Cmca12 structure.
Moreover the CEIMC conductivity of C2c structure is lower than the corresponding PIMD values probably due to the transition to the Cmca-4 structure observed with PIMD. 
It should be noticed that a recent work by Eremets \cite{Eremets2016} claims to have experimentally observed  
the onset of metallic behavior around 350 GPa, close to the values corresponding to $r_s=1.31$ as reported in table \ref{tab:ceimc}.

\begin{figure*}
  \centering
  \includegraphics[width=0.8\textwidth]{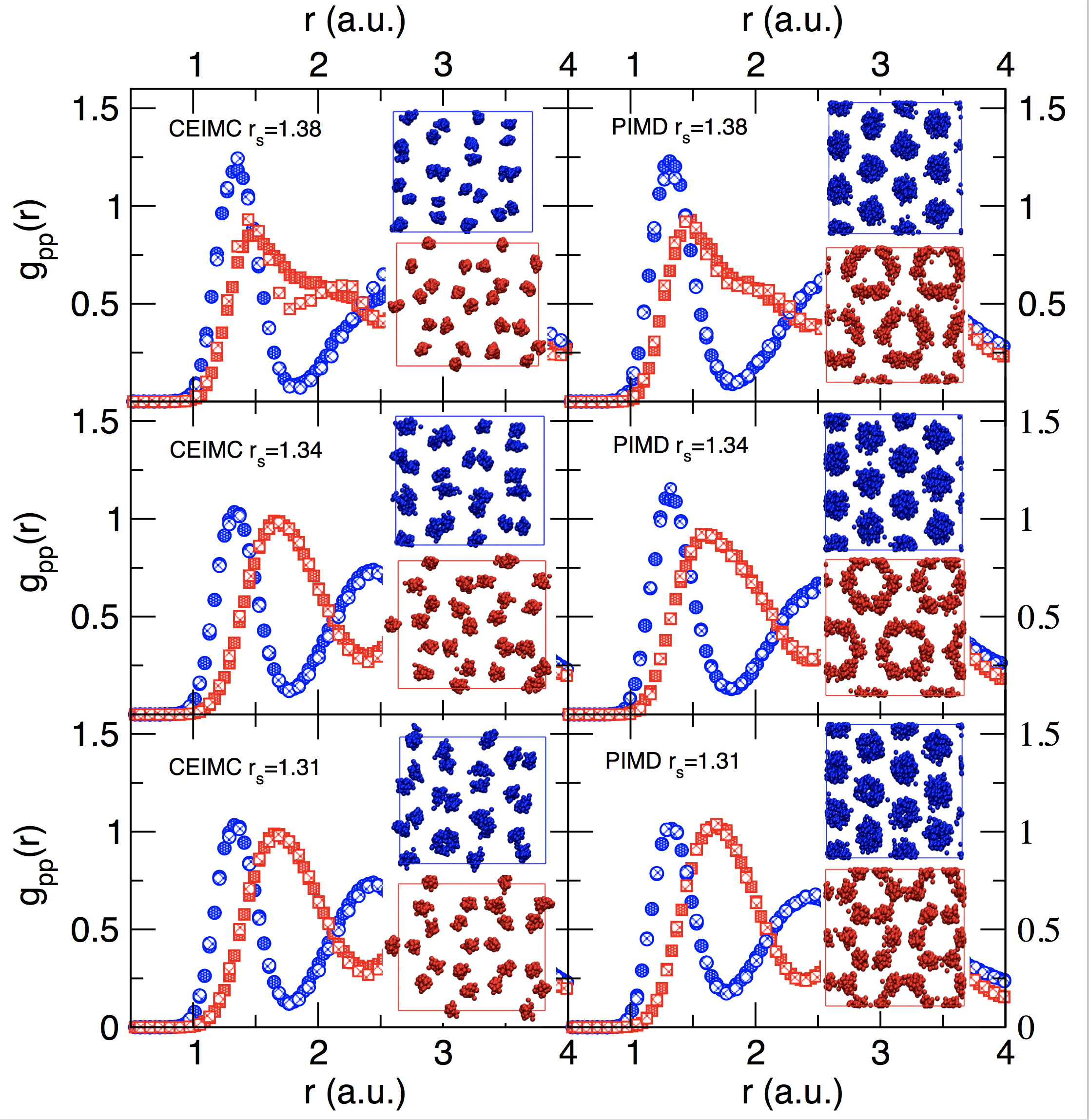}
  \caption{Layer-specific pair correlation functions for CEIMC (left) and PIMD (right) simulations in the Pc48 structure. The densities correspond to pressures from $250 \le P \le 350$ GPa. The four different layers are represented by full and open symbols; layers with strongly bonded molecules are in blue while weakly bonded molecular layers are in red. Insets: snapshots of proton configurations in two adjacent layers.}
  \label{fig:pc48ceimcgr}\end{figure*}

\begin{figure}
  \includegraphics[width=0.4\textwidth]{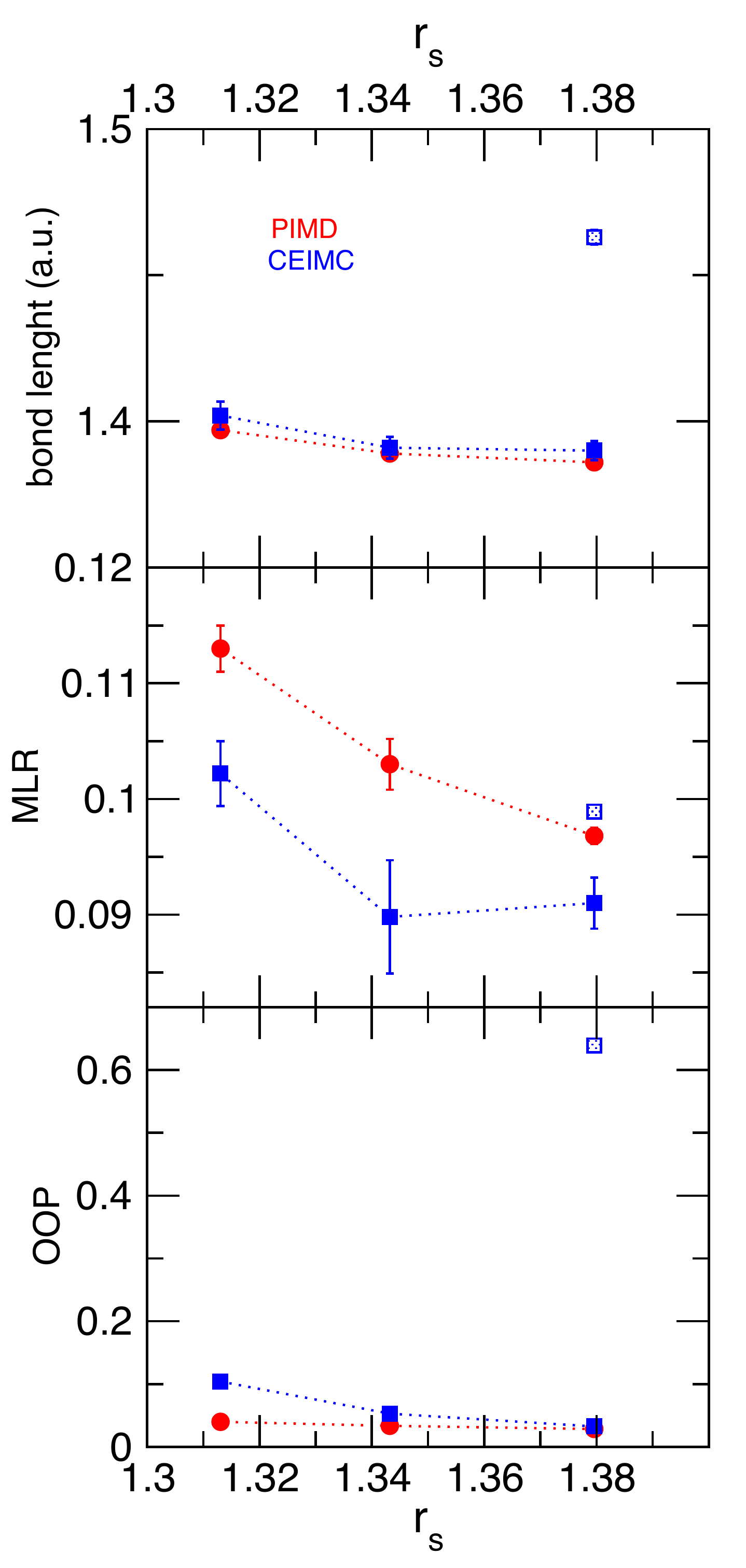}
  \caption{Molecular related structural properties obtained by CEIMC (blue) and PIMD (red) simulations in the Pc48 structure. Upper panel: bond length, middle panel: molecular Lindemann ratio (MLR), lower panel: orientational order parameter (OOP). Closed symbols represent averages the strong molecular layers, while the open symbols (CEIMC at the lowest density) represent averages for the weak molecular layer.}
\label{fig:Pc48struct}
\end{figure}
\begin{figure}
\centering
\includegraphics[width=0.45\textwidth]{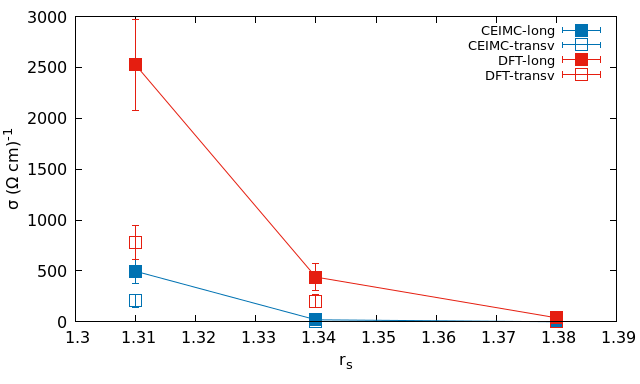}
\caption{Longitudinal and transverse DC conductivity for the Pc48 structure averaged over 16 equilibrium configurations obtained from CEIMC and PIMD simulations.}
\label{fig:Pc48cond}
\end{figure}

\subsection{Pc48 at T=414 K}
We ran both PIMD and CEIMC simulations starting from the relaxed Pc48 lattice, the most stable candidate structure identified
 for phase IV at P=250, 300 and 350 GPa (corresponding to  densities $r_s=1.38,1.34,1.31$ respectively), within DFT using the vdW-DF functional. We performed simulations along the T=414K isotherm, a temperature close to the expected melting line. 

Pc48 is also a layered structure. Thus, we use layer-specific pair correlation functions
to resolve differences among layers, as in the C2c case.
In fig. \ref{fig:pc48ceimcgr} we compare PIMD and CEIMC pair correlation
functions: they show a good agreement, although some of the CEIMC results might still be affected by noise.  
As in the C2c structure, the ``blue'' layers represent the strongly molecular layer while the ``red'' layers are weakly molecular with a broad first peak in which molecules cannot be easily isolated from the rest of the pair correlation function. 
The strong molecular layers (``blue'') do not qualitatively change with increasing density.
The blue insets of the PIMD simulations show strong rotational activity of the 
molecules in those layers; this is not so evident in the CEIMC insets, which seem to display a significant
rotational activity only at higher densities. 
Within both methods, the orientational order parameter of those layers is small as shown in fig. \ref{fig:Pc48struct}. 
In the weak molecular layers (``red''), the main peak gradually looses its initial double--peak structure and moves to higher distances as the density increases. The initial double-peak structure comes from two different probability maxima, the shorter for the intra-ring distance, the longer for the distance between adjacent rings. 
This double--peak structure is seen in both CEIMC and PIMD data at $r_s=1.38$ but only in CEIMC data at $r_s=1.34$ while it is absent at $r_s=1.31$ in both theories. Inspection of the configurations (red insets) shows that this behavior corresponds to the tendency to rearrange the rings of three molecules (six protons) observed at lower density in an hexagonal pattern when density is increased. 
The PIMD simulations display a higher degree of order: for example, well separated rings
can be detected at $r_s=1.34$. An analysis of the PIMD trajectories at this density shows a collective rotation of 6 protons.
These rotations are rare: we could observe 1-2 events in $\approx$ 2 ps, the length of an entire 
PIMD simulation. No rotations were detected in the CEIMC simulations.
We see that the rings made up of three
molecules get closer as the density is increased, resulting in an
increased overlap of charge among atoms in these rings. Eventually at $r_s=1.31$ a symmetric hexagonal structure is obtained.

Fig. \ref{fig:Pc48struct} shows the structural properties for Pc48.
In the ``red'' layer it is difficult to distinguish protons belonging to a specific molecule at the higher densities. For this reason, structural features of molecules in these layers are reported only at lower densities for PIMD and at the 3 lowest densities for CEIMC.

Results for the electrical conductivities are reported in fig.\ref{fig:Pc48cond}. The qualitative features are the same as those for the other structures considered: the band gap shrinks and, at some critical value of $r_s$, becomes zero; the electronic density of states has a depression around the Fermi energy which is filled upon increasing density. 
Results in fig. \ref{fig:Pc48cond} show that the system becomes metallic at $r_s=1.34$  with PIMD while
the metallic state is found only at $r_s=1.31$ (P=331GPa) with CEIMC. Moreover, the conductivity in PIMD is larger, probably because the higher nuclear motion in the weakly bonded molecular layers favors the metallic state. In both cases, however, the conductivity is low when compared with standard metals.

\section{Discussion and conclusions}
\label{sec:discussion}
From a theoretical standpoint, we can  compare our findings with
previous dynamical DFT simulations and static QMC
calculations.
Previous dynamical calculations were mostly performed at higher temperatures (except for ref. \cite{Morales2013}),
showing qualitative features similar to our PIMD results: a stable C2c at lower densities; a first transition to the Ibam structure when increasing pressure;  a new transition to Cmca-4 at higher pressures \cite{Liu2013,Goncharov2013,Magdau2013,Liu2014a}. 
These works mainly focused on the properties of the mixed structures,
especially on their hexagonal layers, finding contradicting results on the nature of the mixed structure and on the possibility
of proton diffusion. However, as pointed out in refs. \cite{Morales2013,Chen2014,Azadi2013b}, the energy
 landscape of high pressure hydrogen is not well represented by semilocal DFT
 functionals.  Nuclear quantum effects enhance the
 inaccuracy. But MD with classical protons might result in a cancellation of errors depending on the chosen functional \cite{Morales2013liquid}.
In the present work, we found that the accuracy of the PIMD results with the vdW-DF approximation depends on the initial
symmetry of the crystal. For Cmca12 symmetry, PIMD simulations give a picture qualitatively 
consistent with CEIMC.However, for the C2c symmetry, the trend of layering transitions with pressure observed with PIMD and ending in the metallic Cmca-4 symmetry is not confirmed by CEIMC. 
Ref. \cite{Chen2014} suggests that energy barriers leading to molecular dissociation are underestimated with DFT; our
 results confirm that 
 molecules do not dissociate at T=200 K up to at least $r_s=1.27$ (P$\approx$ 430 GPa). We see first signs of
 dissociation at $r_s=1.23$ only (P $\approx$ 520 GPa) for C2c at 200K or at $r_s\sim 1.34$ (P$\approx$290GPa) when considering Pc48 at T=414 K. 
For this structure, protons in the weakly bonded layers tend to form rings, which become similar to a hexagonal
network at the highest density considered. Once this symmetric hexagonal structure is reached, the molecular character is lost (the first peak of the g(r) is now at $r\sim 1.75 a_0$) and the system is metallic but having low conductivity.

A precursor of the molecular dissociation is the large protonic motion and the rotation of the hexagonal rings observed in the ``red'' layer with PIMD.
Previous works\cite{Liu2014a}  found that protons could diffuse through the layer.  As discussed above, 
our simulations were $\approx$ 2 ps long, too short to see any diffusion event linked to the rotations of the rings, as pointed out by Liu et al.\cite{Liu2014a}. 
It is possible that protonic diffusion is caused by rotations of different rings: this process is favored at high density, when rings ``merge'' in a  hexagonal network, as we observed at $r_s=1.31$. 
We did not see different kinds of hexagonal layers, as proposed by Magdau et al. \cite{Magdau2013} 
but this was expected because of the limited size of the simulation box.

In PIMD simulations for the C2c structure, we also observed rare concerted rotation of three molecules in the weakly molecular layers at lower densities:
this led to important differences in CEIMC and PIMD predictions for structural properties. This is another indication that energy barriers (in this case, involving rotations) are underestimated in the DFT potential energy surface. 
Some discrepancies between PIMD and CEIMC are found at higher temperature as well: the rotational character of 
the strongly bonded molecules is different; the protonic motion in the weakly bonded layer is enhanced in the PIMD simulations.  

A direct comparison with static QMC calculations is harder: these computations focus on the relative stability of the different candidate structures, that cannot be assessed during dynamical NVT simulations. 
Phase diagrams obtained in refs. \cite{McMinis2015b,Drummond2015} show a large stability region for C2c,
in agreement with our present finding from CEIMC. In contrast, dynamical PIMD predicts spontaneous transitions to other structures (Pbcn, Ibam, Cmca4) in the pressure range corresponding to the densities considered ($P>250$ GPa). 


 In conclusion we performed exploratory PIMD and CEIMC simulations of selected structures 
for solid hydrogen at high pressure along an isotherm at T=200 K  in phase III and an isotherm at 414 K  in phase IV.
The comparison between PIMD and CEIMC at T=200 K shows how the dynamical stability of the C2c structure can be underestimated by DFT. 
This suggests that the energy surface from vdW-DF, the adopted xc functional, does not adequately describe free energy barriers thus allowing
in-plane rotations and favoring structural rearrangements. 
Conversely for the Cmca12 structure, the two theories are in qualitative agreement although quantitative difference are observed.
 Both C2c and Cmca12 structures exhibit a semi-metallic
 behavior starting at $\sim$ 350 GPa, which is consistent with some recent experimental claims \cite{Eremets2016} .
The results of Pc48 at T=414 K show quantitative but not qualitative differences between CEIMC and PIMC.
 
 \begin{acknowledgments}
We would like to thank Markus Holzmann, Yubo Yang and Dominik Domin for useful discussions. M.A.M. was supported through the Predictive Theory and Modeling for Materials and Chemical Science program by the US Department of Energy (DOE) Office of Science, Basic Energy Sciences. This work was performed in part under the auspices of the US DOE by Lawrence Livermore National Laboratory under Contract DE-AC52-07NA27344. D.M.C. was supported by DOE Grant NA DE-NA0001789 and by the Fondation NanoSciences (Grenoble). C.P. was supported by the Agence Nationale de la Recherche (ANR) France, under the program ``Accueil de Chercheurs de Haut Niveau 2015'' project: HyLightExtreme. Computer time was provided by PRACE Projects 2013091918 and 2016143296 and by an allocation of the Blue Waters sustained petascale computing project, supported by the National Science Foundation (Award OCI 07- 25070) and the State of Illinois.  
\end{acknowledgments}

\bibliographystyle{apsrev4-1}

\bibliography{dottorato}

\end{document}